\documentclass{PoS}

\title{Toward a precise determination of  $\mathbf{T_c}$ with 2+1 flavors of quarks}

\ShortTitle{$T_c$ determination}

\author{
  \speaker{Carleton DeTar and Rajan Gupta} 
  (HotQCD Collaboration\thanks{
  Tanmoy Battacharya,
  Michael Cheng,
  Norman Christ,
  Carleton DeTar,
  Steven Gottlieb,
  Rajan Gupta,
  Urs Heller,
  Kay Huebner,
  Chulwoo Jung,
  Frithjof Karsch,
  Edwin Laermann,
  Ludmila Levkova,
  Thomas Luu,
  Robert Mawhinney,
  Peter Petreczky,
  Dwight Renfrew,
  Christian Schmidt,
  Ron Soltz,
  Wolfgang Soeldner,
  Robert Sugar,
  Doug Toussaint,
  and
  Pavlos Vranas
  }
  ) \\
  Physics Department, University of Utah, Salt Lake City, UT 84112, USA\\
  Theoretical Division, Los Alamos National Laboratory, Los Alamos, NM 87545, USA\\
  E-mail: \email{detar@physics.utah.edu}, \email{rajan@lanl.gov}
}

\newcommand{\pbp}{\langle \bar \psi \psi \rangle}

\abstract{We present a status report on a new high statistics study of
  the high temperature transition in full QCD at zero chemical
  potential.  Our simulations use both improved asqtad and p4
  staggered quarks on lattices with a temporal extent $N_\tau = 8$ and
  light quark masses approximately one
  tenth the strange quark mass.  In this report we describe the
  setup of our calculations and present a preliminary analysis of a
  variety of sources of systematic error and ambiguity in the
  determination of the crossover temperature. We propose to present our 
  final analysis with double the current statistics. These calculations were
  carried out on the IBM BlueGene/L supercomputer at 
  Lawrence Livermore National Laboratory.}

\FullConference{The XXV International Symposium on Lattice Field Theory\\
		 July 30 - August 4 2007\\
		 Regensburg, Germany}

\begin{document}

\section{Introduction}

The high temperature transition in strongly interacting matter is
currently being investigated in major experiments at the Brookhaven
relativistic heavy ion collider (RHIC) and will soon be studied at the
LHC as well.  Lattice simulations of QCD provide an {\it
ab initio} characterization of strongly interacting matter in or close
to thermal equilibrium and at or close to zero baryon and strangeness
chemical potential. By quantifying the behavior of the equation
of state, changes in the heavy quark potential and transport
coefficients as a function of temperature, lattice results will
provide crucial guidance in the phenomenological interpretation of
experimental measurements.

The first set of quantities that can be determined precisely
using lattice QCD are the transition temperature, strangeness
susceptibility, and the equation of state.  The importance of a
reasonably accurate determination of the transition temperature is
obvious since the minimum energy densities required to produce a quark-gluon
plasma grow approximately as the fourth power of the temperature.
Thus, a 10\% error in the threshold temperature corresponds to a 45\%
error in the threshold energy density.

Thermodynamic simulations face the standard set of challenges inherent
in all lattice QCD simulation, {\it i.e.} of achieving reliable
continuum and chiral extrapolations.  Our calculations have been
carried out with two sets of ${\cal O}(a^2)$ improved actions --
asqtad and p4 staggered fermions. With these actions an imaginary time
extent $N_\tau = 8$ and $m_\pi \approx 215$ MeV represent the state of
the art. To check the efficacy of our approach we will compare our
results with those of a less improved action at smaller lattice
spacing ($N_\tau = 10$) and lower pion mass \cite{Aoki:2006br}.

Recently three groups have presented results on the transition
temperature. The MILC Collaboration obtains $T_c=169(10)(4)$
Mev~\cite{Bernard:2004je} based on analyzing the chiral susceptibility,
while the RBC-Bielefeld Collaboration reports $T_c=192(7)(4)$, based
on both the chiral susceptibility $\chi_\ell$ and the Polyakov loop
susceptibility $\chi_L$~\cite{Cheng:2006qk}.  The Budapest-Wuppertal
Collaboration carried out simulations closer to the continuum limit
with physical quark masses but uses an action that is not ${\cal
  O}(a^2)$ improved and finds $T_c=151(3)(3)$ MeV, using $\chi_\ell$,
and $T_c= 175(2)(4)$ MeV and $T_c=176(3)(4)$ MeV from the strange
quark number susceptibility $\chi_s$ and the renormalized Polyakov
loop $L_{\rm ren}$, respectively~\cite{Aoki:2006we,Aoki:2006br}.  We
are working towards developing a quantitative understanding of
systematic errors, which were different in the three calculations. Our
goal is to resolve the order $\pm20$ MeV differences in $T_c$ in the
current estimates and reduce the uncertainty to $\sim 5$ MeV.

The Lawrence Livermore National Laboratory houses a 64-rack IBM Blue
Gene/L computer. Each rack, consisting of 1024 nodes, is capable of a
sustained performance of about one TFlops on lattice QCD codes.  Our
collaboration (HotQCD) was formed in 2006 after we secured permission
from the US DOE (NNSA) to carry out QCD thermodynamic simulations on a
portion of this resource.  The high-security location of the computer
creates an unusual computing environment.  Only three of us could
supervise the actual simulation, and a very limited set of
computational results could be brought out -- then only on paper.  The
detailed logs and gauge configuration files are archived for
subsequent analysis but cannot be moved from the secure computing
environment.  To assure the integrity of the results we matched
selected calculations done on identical computers inside and outside
the security perimeter and verified checksums for each printed output
line.

\section{Transition markers}

At zero baryon number and strangeness there is a deconfinement phase
transition in the limit of infinite quark masses and a chiral symmetry
restoring phase transition when the quark masses vanish.  Between
these extremes we may have a phase transition or simply a crossover.
Indeed, it is widely believed that at physical quark masses and zero
baryon number and strangeness, the high temperature transition is a
rapid crossover, rather than a genuine phase transition
\cite{Karsch:2001cy}.
%
%
The crossover exhibits characteristics of both deconfinement and
chiral symmetry restoration.  Different observables are more sensitive
to one or the other characteristic.  For example, the light-quark
chiral condensate $\pbp_\ell$ is an order parameter for the chiral
phase transition at $m_\ell = 0$.  At physical quark masses it is
large at low temperature and drops to a low value over a narrow range
of temperatures as characteristic of approximate chiral symmetry
restoration at high temperature.  An inflection point in this
crossover as a function of temperature is a marker of a transition in
the chiral properties of the medium.  Similarly, the associated
isosinglet chiral susceptibility $\chi_{\rm singlet}$ measures
fluctuations in $\pbp_\ell$, and a peak in $\chi_{\rm singlet}$ also
serves as a marker for the chiral transition.

Among indicators more directly related to deconfinement, the Polyakov
loop $L \propto \exp(-F_Q/T)$ measures the free energy $F_Q$ of a
static quark in the medium.  In the confined phase the static quark is
screened by a bound light quark, and $F_Q$ reflects the relatively 
large binding energy of the light quark.  In the deconfined
phase it can be screened by collective effects in the plasma at
a much lower cost in free energy.  Thus the Polyakov loop rises at the
crossover and its inflection point marks deconfinement.  Furthermore,
as a result of deconfinement one expects a much reduced cost in free
energy to add a light quark or strange quark to the ensemble.  So the
baryon number and strangeness number susceptibilities rise at the
transition and their inflection points also mark deconfinement.

It has long been understood that when there is only a crossover, the
transition temperature determined from different markers need not
agree\footnote{
For a recent discussion, see \protect\cite{Aoki:2006br}.}
-- agreement is expected only at a critical
point.  Our first goal is to quantify the differences at
the physical quark masses by determining each marker precisely. 

For phenomenological applications, some very important
markers are the rapid
rise in energy and entropy density as a function of temperature. Locating the
crossover in the energy density requires an improved determination of
the equation of state to which the above quantities contribute.  However,
for such a determination analogous measurements at $T=0$ are needed.
These are still in more preliminary stages.

\section{Sources of error and ambiguity}

A primary purpose of this study is to make a qualitative assessment of
some of the important sources of error and ambiguity in the
determination of the transition temperature.  We discussed sources of
ambiguity in determining what $T_c$ to associate with a crossover in
the previous section.  Here we list additional sources of statistical and
systematic error.

\begin{enumerate}
  \item Finite volume.  Fluctuations in the light quark chiral order
    parameter are long range and sensitive to the lattice spatial
    volume. 
    \label{item:volume}
  \item	Statistics. The current sample sizes are given in Tables~\ref{tab:p4param}
    and~\ref{tab:asqtadparam}.
  \item Locating the position of the peak or inflection point.  The determination of the
    location of a peak or inflection point inherits the statistical
    errors in the observable itself.  Compared with peaks, inflection
    points are determined by the vanishing of a higher order
    derivative, so require a higher level of accuracy in the data. 
  \item Extrapolation to physical quark masses and the continuum limit.  To obtain
    the transition temperature at the physical point, one must make
    measurements at a variety of quark masses and lattice spacings and 
    extrapolate. 
  \item Error in the determination of the lattice scale.  The
    scale in the Budapest-Wuppertal work is set through separate measurements 
    of the kaon decay
    constant at zero temperature $f_K$ \cite{Aoki:2006br}.  We, 
    at present, determine it through a separate
    measurement of the static quark potential $V(r)$ at zero
    temperature with the same lattice parameters.  The parameters
    $r_0$ and $r_1$, defined through $r^2V^\prime(r)|_{r = r_0} = 1.65$
    and $r^2V^\prime(r)|_{r = r_1} = 1$, are, in turn, determined in
    physical units from independent measurements of $\Upsilon$
    splittings.  Thus the values of $r_0 = 0.469(7)$ fm and $r_1 =
    0.318(7)$ fm are known to $\sim 2$ percent \cite{Aubin:2004wf}.
    At nonzero lattice spacing, different methods
    for determining the scale can disagree, but all methods must agree
    up to statistical errors in the continuum limit.
  \item R-algorithm step size error.  Earlier calculations done with the R
    algorithm suffered from a systematic error introduced by a nonzero
    molecular dynamics step size $dt$.  The more recent RHMC algorithm
    is exact.  We will eventually be combining our current RHMC
    results with earlier R algorithm calculations. We have, therefore, 
    performed simulations with identical parameter sets to estimate 
    the uncertainty associated with the step size choices in earlier 
    R algorithm calculations.
    \label{item:step}
\end{enumerate}

In this preliminary study we have results bearing on points
\ref{item:volume} and \ref{item:step} in this list and some discussion
of $O(a^2)$ errors.

\section{Parameter set}

The study was carried out on $32^3\times 8$ lattices at a bare quark
mass ratio $m_\ell/m_s = 0.1$ along lines of approximately constant
physics.  That is, the bare strange quark mass was adjusted along the
trajectory to produce an approximately constant physical value of
$m_{\bar ss} = 686$ MeV.  Resulting parameters for the p4 action simulation are
given in Table~\ref{tab:p4param}.  Parameters for the asqtad action
are given in Table~\ref{tab:asqtadparam} and were fixed following the
previous R algorithm study, which, in turn were set from parameters
used in early ensemble production \cite{Bernard:2001av}.  The
resulting strange quark mass along the asqtad trajectory is now known
to be approximately 20\% higher than the physical strange quark mass.

\begin{table}[ht]
  \begin{center}
\begin{tabular}{llll}
$\beta=6/g^2$  & $am_\ell$ & $T$ MeV &  Total\ \#\ of \\
               &        &        &  Trajectories\\
\hline\hline
$3.460$  &  $0.00313$  &  $154$    &  $10000  $ \\
$3.490$	 &  $0.00290$  &  $169$    &  $10000  $ \\
$3.510$	 &  $0.00259$  &  $179$    &  $11280  $ \\
$3.540$	 &  $0.00240$  &  $194$    &  $11440  $ \\
$3.570$	 &  $0.00212$  &  $209$    &  $12460  $ \\
$3.600$	 &  $0.00192$  &  $225$    &  $11790  $ \\
$3.630$	 &  $0.00170$  &  $241$    &  $12070  $ \\
$3.660$	 &  $0.00170$  &  $256$    &  $11190  $ \\
$3.690$	 &  $0.00150$  &  $271$    &  $10760  $ \\
$3.760$	 &  $0.00139$  &  $313$    &  $10920  $ \\
\hline	                                       	 
$3.525$ &  $0.00240$  &  $186$    &  6510, 6120   \\
$3.530$ &  $0.00240$  &  $189$    &  5450, 5530   \\
$3.535$ &  $0.00240$  &  $191$    &  5140, 5750   \\
$3.540$ &  $0.00240$  &  $194$    &  5410, 6260   \\
$3.545$ &  $0.00240$  &  $196$    &  6280, 6250   \\
$3.550$ &  $0.00240$  &  $199$    &  5790, 6520   \\
\end{tabular}
  \end{center}
\caption{Simulation parameters for the p4 action. For thermalization
  800 trajectories are discarded. The last six were done in two
  separate streams with trajectory counts indicated.
   \label{tab:p4param}
}
\end{table}
\begin{table}[ht]
  \begin{center}
\begin{tabular}{llll}
$\beta=6/g^2$  & $am_\ell$    & $T$ MeV  & Total\ \#\ of \\
               &           &          & Trajectories\\
\hline\hline
$6.4580$      & $0.00820 $ & $141.4$  &  $12965 $ \\  
$6.5000$      & $0.00765 $ & $149.6$  &  $12990 $ \\ 		  
$6.5500$      & $0.00705 $ & $159.8$  &  $12720 $ \\ 		  
$6.6000$      & $0.00650 $ & $170.2$  &  $12405 $ \\ 		  
$6.6625$      & $0.00624 $ & $175.6$  &  $12365 $ \\ 		  
$6.6500$      & $0.00599 $ & $181.0$  &  $12445 $ \\ 		  
$6.6675$      & $0.00575 $ & $186.5$  &  $12660 $ \\ 		  
$6.7000$      & $0.00552 $ & $192.1$  &  $12320 $ \\ 		  
$6.7600$      & $0.00500 $ & $206.0$  &  $12530 $ \\ 		  
$6.8000$      & $0.00471 $ & $215.4$  &  $12195 $ \\ 		  
$6.8500$      & $0.00437 $ & $227.4$  &  $12405 $ \\ 		  
$6.9000$      & $0.00407 $ & $239.8$  &  $12470 $ \\ 		  
$6.9500$      & $0.00380 $ & $252.5$  &  $12625 $ \\ 		  
$7.0000$      & $0.00355 $ & $265.6$  &  $12595 $ \\ 		  
$7.0800$      & $0.00310 $ & $287.6$  &  $12790 $ \\  
\end{tabular}
  \end{center}
\caption{Simulation parameters for the asqtad action. Between
1000-1200 are discarded for thermalization.
  \label{tab:asqtadparam}
}
\end{table}

\section{Results}

All results shown here are preliminary.

\subsection{Approximate order parameters}

\begin{figure}[ht]
  \centerline{
    \includegraphics[width=.5\textwidth]{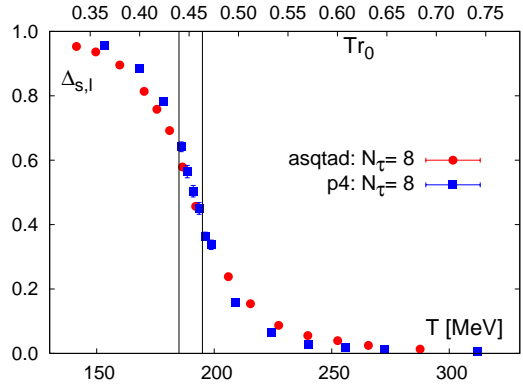}
  }
  \caption{Preliminary results for $\Delta(T)$ for the p4 and asqtad
   actions as a function of temperature in MeV and in units of $r_0$.
   The vertical lines here and throughout mark the range $185-195$
   MeV for purposes of comparison only. 
  \label{fig:delta}
  }
\end{figure}

The unrenormalized chiral condensate and Polyakov loop suffer from
ultraviolet divergences.  The chiral condensate $\pbp_{\ell,s}$ has an
ultraviolet perturbative contribution of the form $m/(a^2)$ at
nonzero quark mass, and the static quark free energy $F_Q$ has a
self-energy divergence of the form $c/a$, contributing a factor
$\exp(-c/aT)$ to the bare Polyakov loop $L_{\rm bare}$.  We consider
the ``subtracted'' condensate \cite{ref:RBC_EOS_2007} and the
``renormalized'' Polyakov loop that are free of
these divergences:
\begin{eqnarray}
  \Delta(T)      &=& \frac{\pbp_\ell(T) - m_\ell/m_s \pbp_s(T)}
                          {\pbp_\ell(0) - m_\ell/m_s \pbp_s(0)} \\
       L_{\rm ren} &=&  \exp[-F_{\infty}(T)/(2T)] \, .\nonumber
\end{eqnarray}
where $F_{\infty}(T)$ is the renormalized free energy of static quark
anti-quark pair separated by infinite distance
\cite{Kaczmarek:2002mc,ref:RBC_EOS_2007}.  The subtracted condensate
is shown for both actions as a function of $T$ in
Fig.~\ref{fig:delta}.  We note that despite slight differences in the
lattice parameters, which would lead to different additive and
multiplicative renormalization factors, these differences largely
cancel in the $ \Delta (T)$ ratio as demonstrated by the reasonable
agreement between the two actions.

\begin{figure}[ht]
  \centerline{
    \includegraphics[width=.5\textwidth]{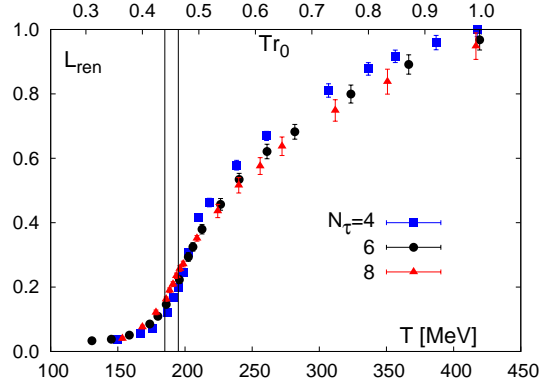}
  }
  \caption{Comparison of the renormalized Polyakov loop data for the
    p4 action with $N_\tau =4$ and $6$ from
    \protect\cite{ref:RBC_EOS_2007} and $N_\tau = 8$ (preliminary from
    this work) as a function of temperature in MeV and in units of
    $r_0$.
  \label{fig:Ploop}
  }
\end{figure}

The renormalized Polyakov loop, shown in Fig.~\ref{fig:Ploop}, is
currently available only for the p4 action.  Comparing ($N_\tau = 4$,
$6$ and 8) data it is evident that $O(a^2)$ scaling violations are
small and the determination of the inflection point has an uncertainty
of a few MeV.
\subsection{Chiral symmetry restoration}

The isosinglet chiral susceptibility measures fluctuations in the
light quark condensate. It is defined as
\begin{eqnarray}
  \chi_s &=& \chi_{\rm dis} + 2 \chi_{\rm con} \\ \nonumber
  \chi_{\rm dis} &=&\left\langle \int d^4r \langle \bar \psi(r) \psi(r) \rangle
          \langle \bar \psi(0) \psi(0) \rangle \right\rangle 
          - V \bigg\langle \pbp \bigg\rangle^2\\
  \chi_{\rm con} &=& \left\langle\int d^4r \langle \bar \psi(r) \psi(0) \rangle
       \langle \bar \psi(r) \psi(0)\rangle\right\rangle \nonumber
\end{eqnarray}
Where $\pbp$ includes both up and down flavors and the ``connected''
and ``disconnected'' subscripts refer to the two kinds of valence
quark line contractions in the correlator.  Here the inner angle
brackets indicate the quark propagator on a single gauge configuration,
and the outer angle brackets denote an average over gauge
configurations.

\begin{figure}[ht]
  \begin{tabular}{cc}
    \includegraphics[width=.5\textwidth]{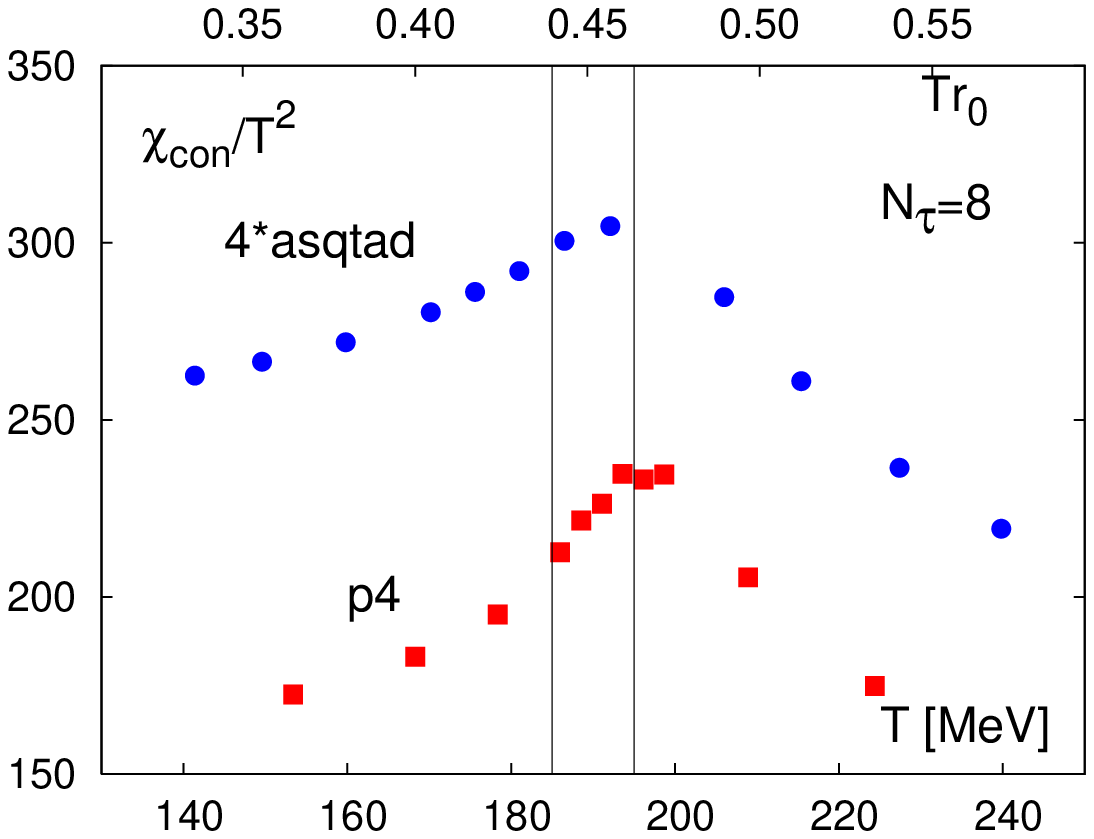}
    &
    \includegraphics[width=.5\textwidth]{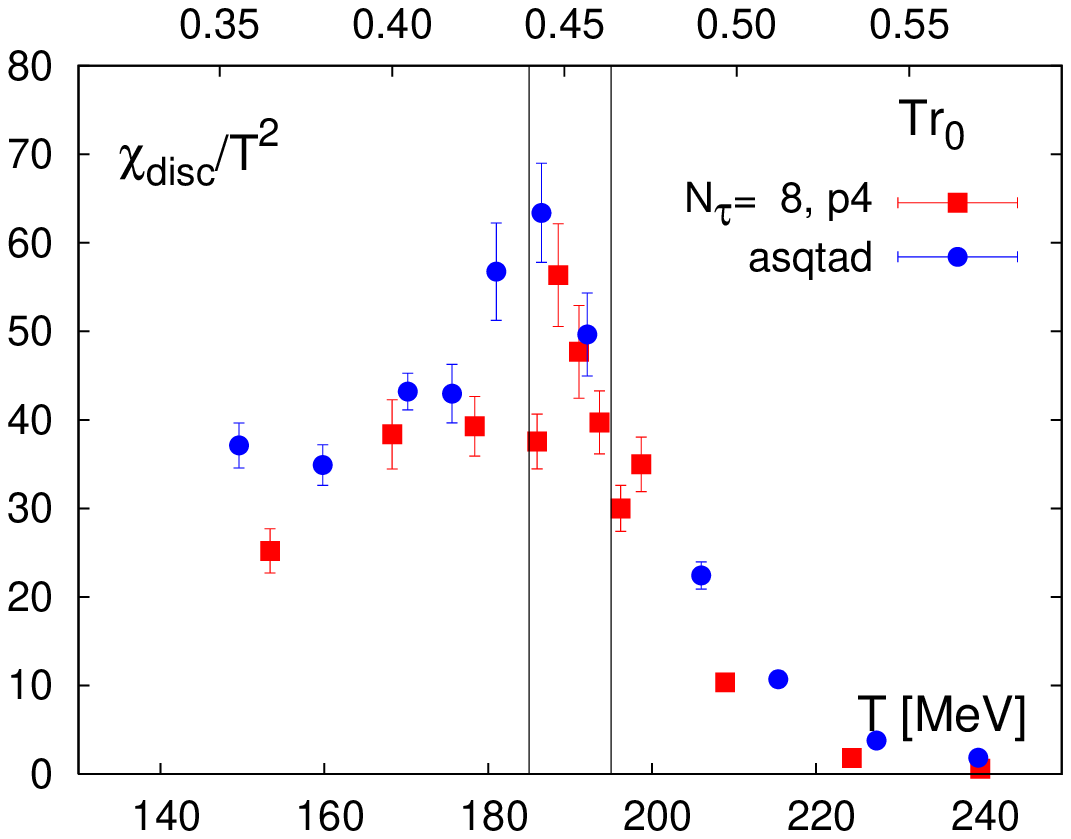} \\
  \end{tabular}
  \caption{Preliminary results for the connected and disconnected
    light quark chiral susceptibilities for both p4 and asqtad action.
    An arbitrary scale factor has been applied to the asqtad connected
    susceptibility to facilitate the comparison.
  \label{fig:p4asqtadChiDisConn}
  }
\end{figure}

\begin{figure}[ht]
  \centerline{
    \includegraphics[width=.5\textwidth]{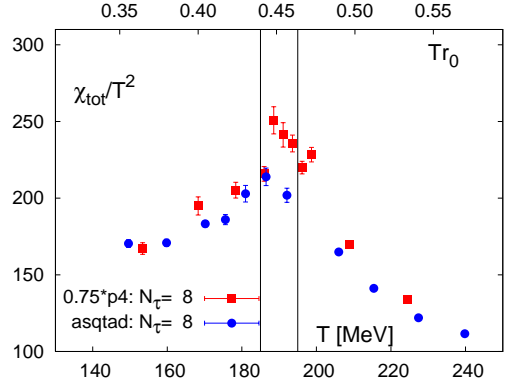}
  }
  \caption{Isosinglet chiral susceptibility for both actions
    (preliminary).  An arbitrary scale factor has been applied to the
    asqtad value to facilitate the comparison.
  \label{fig:p4asqtadChiSing}
  }
\end{figure}

The connected and disconnected light-quark chiral susceptibilities for
the two actions are compared in Fig.~\ref{fig:p4asqtadChiDisConn} and
the total isosinglet chiral susceptibility is shown in
Fig.~\ref{fig:p4asqtadChiSing}.  Since we are now comparing
unrenormalized quantities, differences in normalization are not
surprising.
We are in the process of doubling the statistical sample to resolve
some of the discrepancies evident here.  Nevertheless, it is unlikely
that the location of the peak will shift significantly from the range
$185-195$ MeV at these quark masses and $N_\tau = 8$.

\begin{figure}[ht]
  \centerline{
    \includegraphics[width=.45\textwidth]{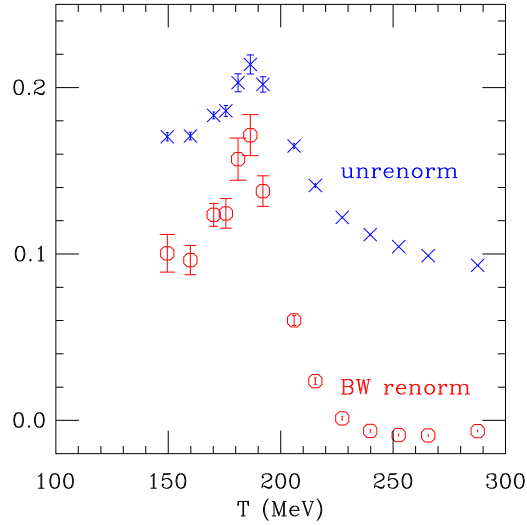}
  }
  \caption{Preliminary unrenormalized isosinglet chiral susceptibility
    for the asqtad action compared with the proposed
    Budapest-Wuppertal renormalized quantity.  To facilitate the
    comparison the unrenormalized susceptibility has been
    rescaled by a factor of 1000.
  \label{fig:asqtadBWChiSing}
  }
\end{figure}

To compensate for additive and multiplicative renormalization factors
in the chiral susceptibility, the Budapest-Wuppertal group has
proposed the modified quantity
\begin{equation}
  m_\ell^2[\chi_{\rm singlet}(T) - \chi_{\rm singlet}(0)]/T^4.
\end{equation}
For the asqtad action we compare this with the unrenormalized
quantity in Fig.~\ref{fig:asqtadBWChiSing}.  To compute this modified
quantity we have used measurements on a small set of zero temperature
($32^4$ lattice) runs with parameters matched to the $N_\tau = 8$
ensembles.  Since the $T=0$ data are smooth, dividing by the
temperature in this way necessarily shifts the peak towards smaller
$T$, however, the shift appears to be only at the level of a few MeV.
\begin{figure}[ht]
  \begin{tabular}{cc}
    \includegraphics[width=.5\textwidth]{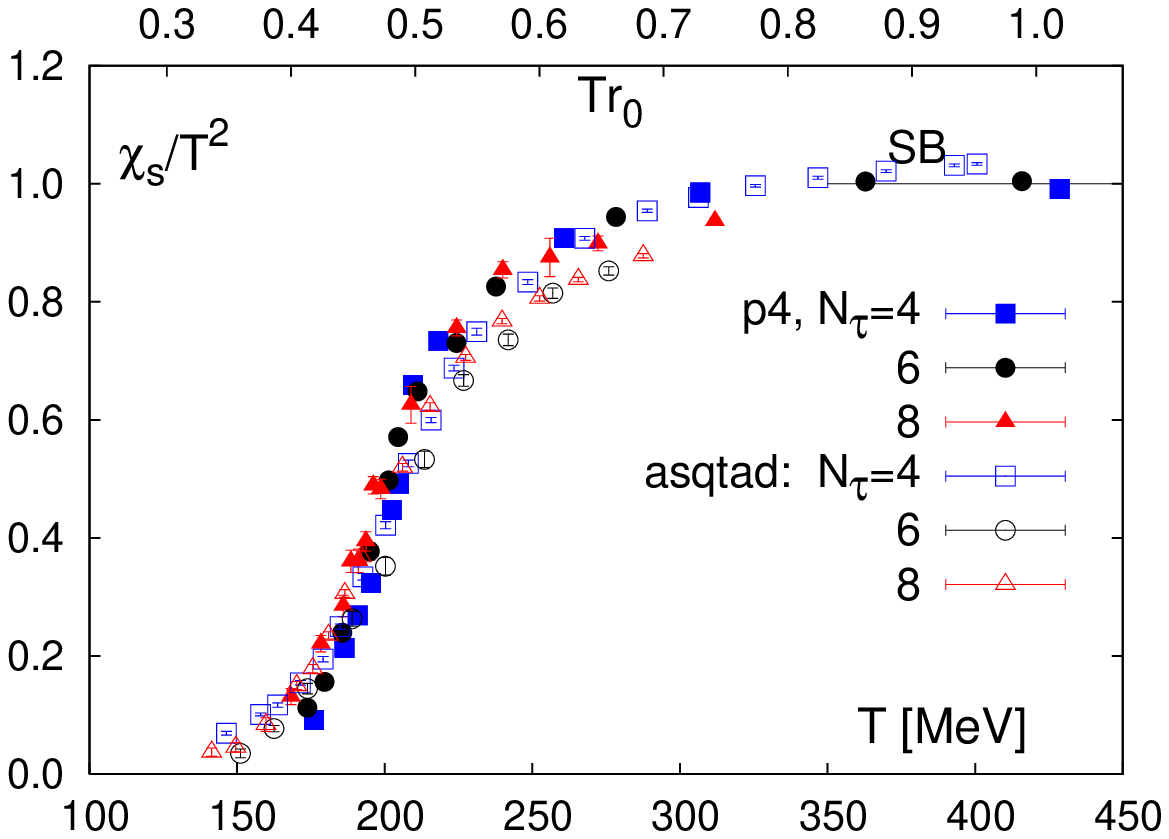}
   &
    \includegraphics[width=.29\textwidth]{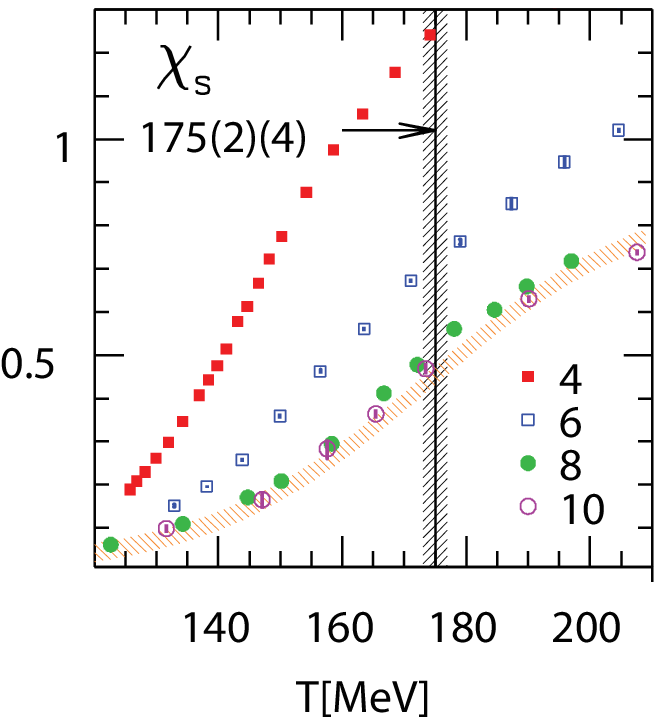}
   \\
  \end{tabular}
  \caption{Left panel: preliminary strange quark number susceptibility
    for the p4 and asqtad actions along the $m_\ell = 0.1m_s$ line of
    constant physics.  Points for increasing $N_\tau$ show scaling
    trends.  Right panel: the same quantity in the crossover region
    for the one-link stout action of the Budapest-Wuppertal
    Collaboration with light quark masses approximately at their
    physical values \protect\cite{Aoki:2006br}.  Somewhat different
    conventions were used to set the temperature scales in the two
    figures.
  \label{fig:p4asqtadChiS}
  }
\end{figure}

\subsection{Crossover in probes of deconfinement}

The light quark number and strange quark number susceptibilities
measure fluctuations in baryon number and strangeness.  They are
defined as
\begin{equation}
  \chi_{(\ell,s)}/T^2 = 
   \frac{1}{VT}\frac{\partial^2 \log Z}{\partial \mu^2_{(\ell,s)}}.
\end{equation}
Since these susceptibilities measure charge, there are no
renormalization issues.  We compare the strange quark number
susceptibility for the two actions in Fig.~\ref{fig:p4asqtadChiS} as
the data are less noisy than those for the light quark. We find good
agreement between the two actions.  Also shown for comparison are
results from the one-link stout action of the Budapest-Wuppertal
collaboration.  In that case the lines of constant physics are close
to the physical quark masses.  The similarities and differences in
scaling behavior, {\i.e.\ ${\cal O}(a^2)$ artifacts, of the three
actions are evident.
\begin{figure}[ht]
  \centerline{
    \includegraphics[width=.45\textwidth]{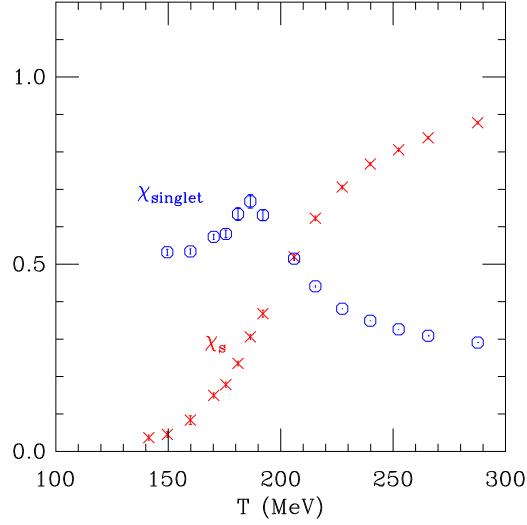}
  }
  \caption{Preliminary results for the strange quark number
    susceptibility for the asqtad action compared with the isosinglet
    susceptibility.
  \label{fig:asqtadChiSingChiS}
  }
\end{figure}

\subsection{Comparison of $T_c$ from deconfinement and chiral symmetry restoration markers}

A central question that this study addresses is how different is $T_c$ 
determined from chiral and deconfinement markers? The data are presented 
in Fig.~\ref{fig:asqtadChiSingChiS}, and we can make the following 
observations. 
\begin{enumerate}
\item
Locating the inflection point in the strange quark susceptibility is
more uncertain than locating the peak in the isosinglet
susceptibility. Estimates of the inflection point are sensitive to the
number of points in the fit and the functional form of the fit.
\item
The peak in the isosinglet chiral susceptibility appears to occur at 
approximately the same temperature as the inflection point in strange
quark number susceptibility.  Further statistics are needed to detect
and quantify any differences.
\end{enumerate}

\begin{figure}[ht]
  \begin{tabular}{cc}
    \includegraphics[width=.45\textwidth]{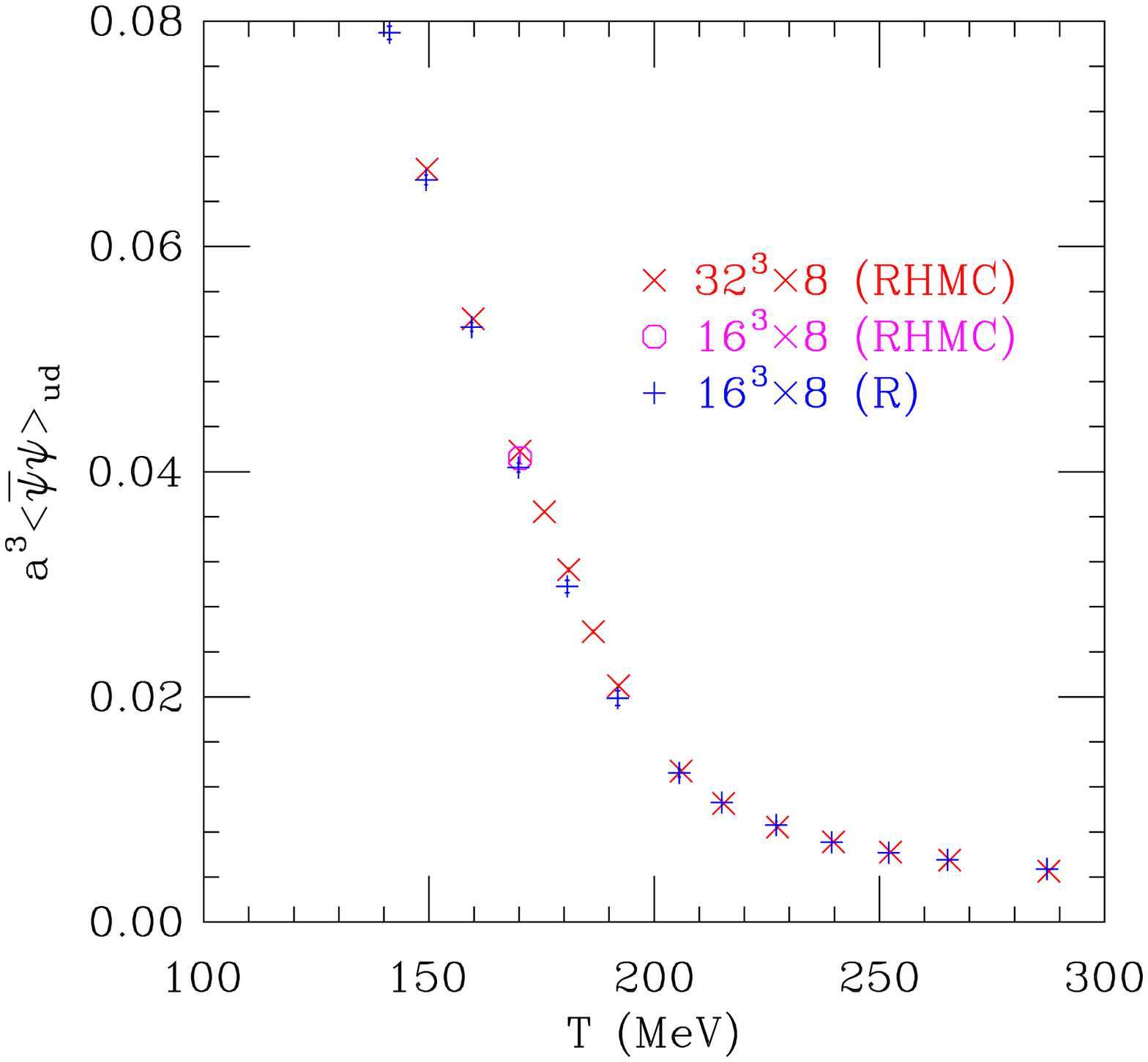}
    &
    \includegraphics[width=.45\textwidth]{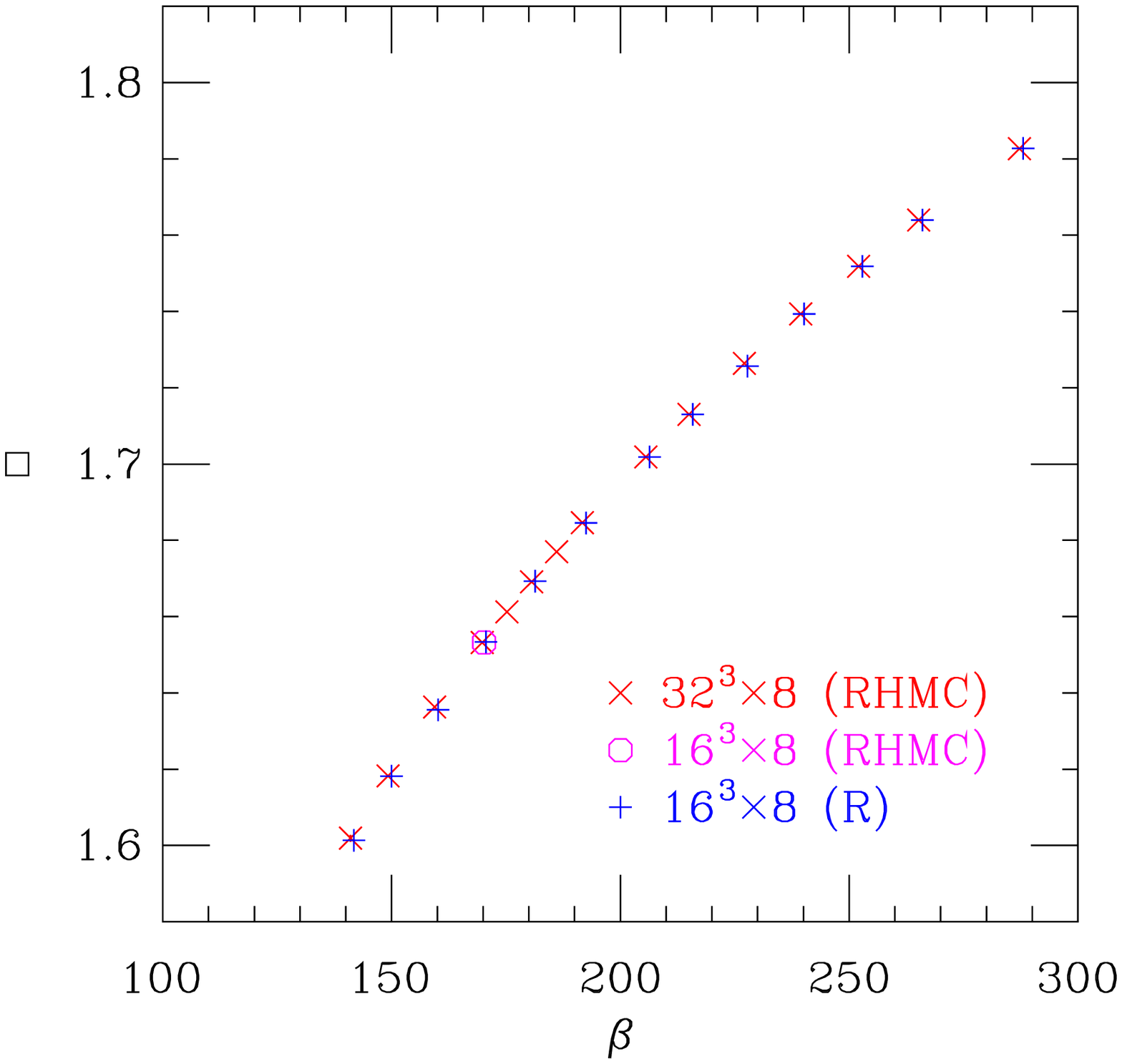}
   \\
  \end{tabular}
  \caption{Light quark chiral condensate and plaquette for the asqtad
    action, comparing results obtained from the R algorithm on
    $16^3\times 8$ lattices and the RHMC algorithm (preliminary
    values) on $32^3\times 8$.  Points for the Polyakov loop have been
    slightly displaced horizontally for clarity.  The $16^3\times 8$
    RHMC point in each case tests sensitivity to the finite spatial size.
  \label{fig:asqtadPbpRePRvsRHMC}
  }
\end{figure}

\subsection{Step size in R algorithm and finite size effects}

We would eventually like to combine new and old results at a variety
of quark masses and lattice spacings, so we can carry out an
extrapolation to the physical point.  Many of the older simulations
used the R algorithm and were done at a smaller aspect ratio
$N_s/N_\tau$.  In Fig.~\ref{fig:asqtadPbpRePRvsRHMC} we compare results
for the two order parameters from simulations with the R algorithm on
$16^3\times 8$ lattices and the RHMC algorithm on $32^3\times 8$.
Differences are very slight.  Although the effect is too small to show
in these figures, there are differences at the level of a little more
than one standard deviation in the chiral condensate at low
temperature but no visible trends in the Polyakov loop.
\begin{figure}[ht]
  \begin{tabular}{cc} 
    \includegraphics[width=.45\textwidth]{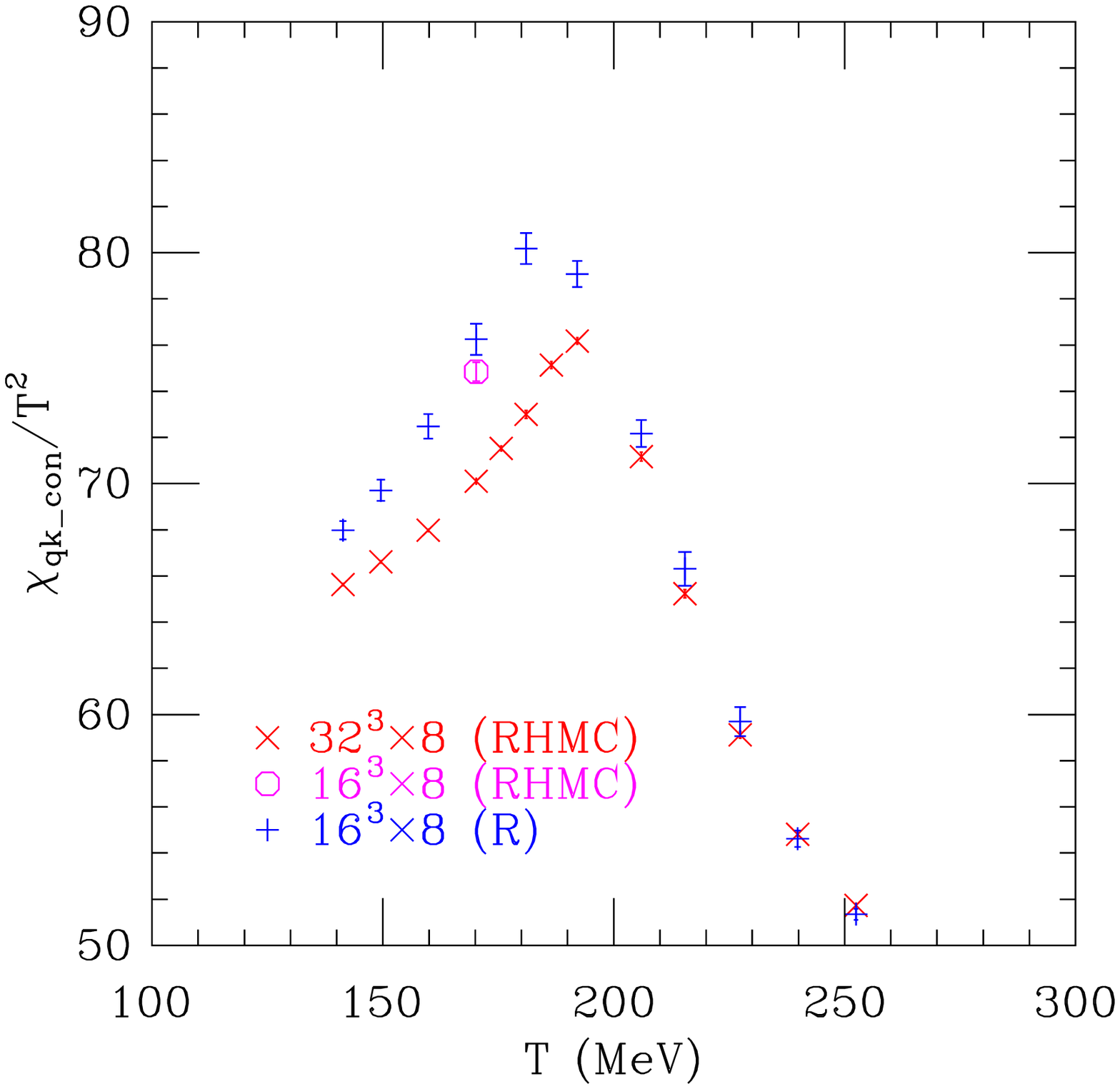}
    &
    \includegraphics[width=.45\textwidth]{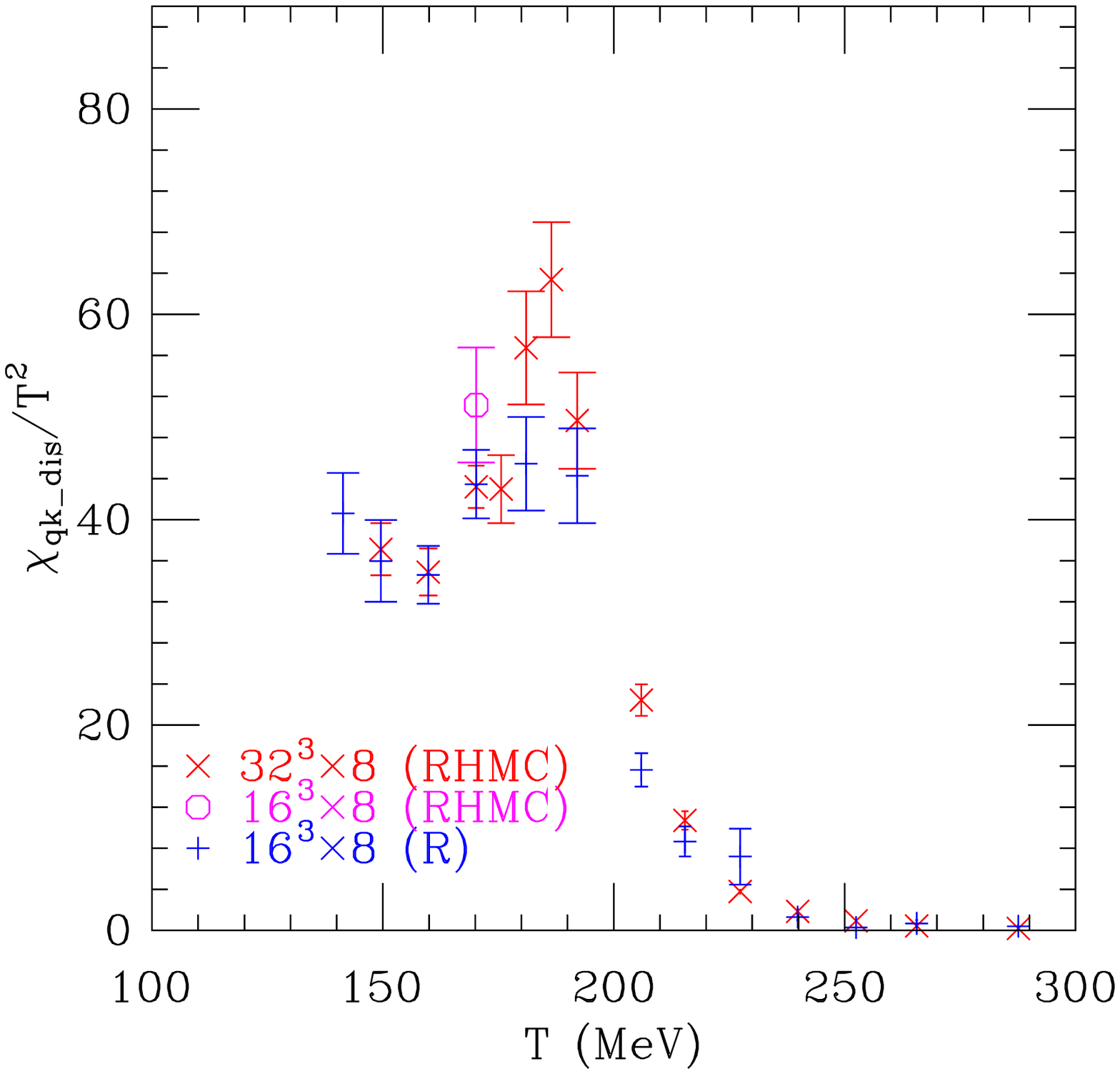}
   \\
  \end{tabular}
  \caption{Light quark chiral connected and disconnected
    susceptibility for the asqtad action, comparing results obtained
    from the R algorithm on $16^3\times 8$ lattices, the RHMC
    algorithm (preliminary results) on $32^3\times 8$, and one point
    with the RHMC algorithm on $16^3\times 8$.
  \label{fig:asqtadChiConnDiscRvsRHMC}
  }
\end{figure}

The light quark susceptibility is more sensitive to differences in the
two asqtad calculations, as shown in
Fig.~\ref{fig:asqtadChiConnDiscRvsRHMC}.  Here the connected
susceptibility is significantly higher at low temperature in the
smaller-volume R-algorithm simulation.  To test whether the effect is
due to the smaller volume or the algorithm we carried out a simulation
of the RHMC algorithm on the smaller volume at one temperature as
shown.  It is close to the R algorithm result.  Thus we conclude that
difference in the connected susceptibility is largely a finite volume
effect.  We also see that the disconnected term appears to have a
sharper peak at larger volume.

When the two susceptibilities are combined to form the isosinglet
susceptibility, the finite volume effect tends to shift the peak
towards higher $T$ by a few MeV with this increase in volume.

\subsection{Conclusions and plans}

We find good agreement between the p4 and asqtad actions, bearing in
mind the small differences in parameter choices.  A preliminary
assessment of the various crossover markers at $N_\tau = 8$ and $m_\ell =
0.1 m_s$ suggests that they may disagree at the level of several MeV,
but not at the level of a few tens of MeV.  We see similar differences
in the peak in the isosinglet chiral susceptibility between the 
unnormalized and Budapest-Wuppertal-normalized version.  We
compared asqtad results from the R algorithm and RHMC and do not
find evidence of significant step-size effects that would affect the
determination of the transition temperature, but we do observe a
finite-volume effect that could lower the temperature of the peak in the
chiral susceptibility by a few MeV.

Overall, we find the crossover in both the deconfinement and chiral
symmetry restoration markers to lie in the range $T=185-195$ MeV at
$N_\tau = 8$ and at $m_\ell = 0.1 m_s$. Our plans for the immediate
future are to double the statistics in the transition region, add data
at more values of the quark masses and carry out a detailed
quantitative analysis of the effects observed here.

\acknowledgments
We are grateful to both LLNL and the NNSA for providing access to the
LLNL Bluegene/L computer.  This work is supported by grants from the
US Department of Energy and US National Science Foundation.

\end{document}